\documentclass[notitlepage,aip,jcp,amsmath,amsfonts,amssymb,floatfix,superscriptaddress,10pt,reprint,longbibliography]{revtex4-2}
\usepackage{color}
\usepackage{xcolor}
\usepackage{braket}
\usepackage{mathtools}
\usepackage{dcolumn} 
\usepackage{graphicx}
\usepackage[breaklinks=true]{hyperref}
\usepackage{ulem}
\usepackage{cancel}

\newcommand{\BTS}{\ensuremath{\mathrm{BTS}}}
\newcommand{\AGP}{\ensuremath{\mathrm{AGP}}}

\begin{document}
\title{Unrestricted Binary Tree State for the Fermionic Hubbard Model}
\author{Mahroo Mirabdolhagh Hazaveh}
\email[Author to whom correspondence should be addressed: ]{Mahroo.Mirabdolhagh.Hazaveh@rice.edu}
\affiliation{Department of Chemistry, Rice University, Houston, TX 77005-1892, USA}

\author{Armin Khamoshi}
\affiliation{Department of Physics and Astronomy, Rice University, Houston, TX 77005-1892, USA}

\author{Fei Gao}
\affiliation{Department of Physics and Astronomy, Rice University, Houston, TX 77005-1892, USA}

\author{Thomas M. Henderson}
\affiliation{Department of Chemistry, Rice University, Houston, TX 77005-1892, USA}
\affiliation{Department of Physics and Astronomy, Rice University, Houston, TX 77005-1892, USA}

\author{Gustavo E. Scuseria}
\affiliation{Department of Chemistry, Rice University, Houston, TX 77005-1892, USA}
\affiliation{Department of Physics and Astronomy, Rice University, Houston, TX 77005-1892, USA}
\date{\today}

\begin{abstract}
Binary tree states have been introduced as compact correlated pair wave functions that improve upon the antisymmetrized geminal power by adding variational flexibility.
Here, we extend the binary tree state to the fermionic half-filled Hubbard model using restricted and unrestricted orbitals.
Although the ansatz is seniority zero in its optimized pairing basis, its expansion in the original Hubbard site basis can contain nonzero site-seniority character, while unrestricted orbital optimization provides additional flexibility through independent spin-up and spin-down orbitals.
This provides a mechanism for a paired ansatz to describe the singly occupied configurations that dominate the repulsive Hubbard model at large $U/t$.
We find that unrestricted BTS closely approaches optimized DOCI in the
repulsive regime, substantially improves upon AGP, and reproduces DOCI
density correlations more accurately than pCCD.
\end{abstract}

\maketitle

\section{Introduction}
 
Strong correlation is often most naturally described in terms of electron pairs.\cite{Racah1943,RingSchuck1980,Surjan1989} 
When bonds are stretched, electrons are localized, or several occupations become nearly degenerate, the wave function must describe more than a single determinant. 
A useful way to express this pairing structure is through seniority,\cite{Bytautas2011,Johnson2013CTC,CaleroOsorio2025} 
which for a fixed orbital basis and pairing scheme counts the number of electrons that are not paired with their assigned partners. 
The seniority-zero space is therefore the subspace in which every electron belongs to a pair. 
Although the definition depends on the chosen orbitals, an optimized pairing basis can make this sector a compact and physically meaningful part of the wave function for many strongly correlated systems.\cite{Bytautas2011,Johnson2013CTC,CaleroOsorio2025}

The exact wave function within the seniority-zero space is referred to as doubly occupied configuration interaction (DOCI).\cite{WeinholdWilson1967,VeillardClementi1967,CoutyHall1997,Bytautas2011} The DOCI Hilbert space is much smaller than that of full configuration interaction, but its cost is still combinatorial. For a system with $m$ orbitals and $n$ electron pairs, the number of DOCI coefficients is $\binom{m}{n}$, roughly the square root of the size of the corresponding full CI space, but the scaling still quickly becomes prohibitive as the system grows.\cite{Bytautas2011,Henderson2014}

This cost has motivated the development of approximate pair-based wave functions for strongly-correlated systems. A formally flexible example is APIG, the Antisymmetrized Product of Interacting Geminals.\cite{Silver1969,Johnson2013CTC} In practice, however, APIG is difficult to use because its coefficients involve permanents.\cite{Silver1969,Johnson2013CTC} More practical approaches impose additional structure. One important example is AP1roG, often referred to as pair coupled-cluster doubles (pCCD).\cite{Limacher2013,Boguslawski2014,Stein2014,Henderson2014} With optimized orbitals, pCCD often reproduces DOCI energies very well in many repulsive strongly correlated problems at low polynomial cost.\cite{Limacher2013,Stein2014,Henderson2014} At the same time, it has well-known weaknesses. It is nonvariational, which makes orbital optimization more delicate.\cite{Henderson2014,Henderson2015} It can also break down in attractive pairing regimes,\cite{HendersonDukelsky2014,Dukelsky2003} and in strongly correlated repulsive systems, broken-pair or higher-seniority effects may become important.\cite{WahlenStrothman2018} In addition, pCCD usually reproduces energies more faithfully than density matrices, and adding dynamical correlation on top of it is possible but not always straightforward.\cite{Henderson2014,Henderson2015,TecmerBoguslawski2022, NowakBoguslawski2023, Garza2015PCCP, BoguslawskiAyers2015, Garza2015JCP, Limacher2014PCCP}

Another important seniority-zero ansatz is the antisymmetrized geminal power (AGP).\cite{Coleman1965,BratozDurand1965,Linderberg1980} In its natural orbital basis, AGP is equivalent to a number-projected BCS wave function.\cite{RingSchuck1980,BlaizotRipka,Schmid,PQT,Khamoshi2023QST,Khamoshi2019} AGP is compact, variational, and computationally efficient, and it often provides a useful qualitative picture of pairing. Its main limitation is that all pairs are built from the same geminal, which makes the ansatz too restrictive for many quantitative applications. For that reason, AGP is often better viewed as a useful variational starting point than as a generally sufficient seniority-zero wave function on its own.\cite{Neuscamman2013,Zen2014,HendersonScuseria2019,HendersonScuseria2020,Dutta2020,Khamoshi2021,KhamoshiEvangelista2021,Dutta2021,Khamoshi2023QST} 

The binary tree state (BTS) is another seniority-zero ansatz which is a more flexible variational correlated-pair ansatz that contains AGP as a special case.\cite{Dutta2024BTS} BTS keeps a compact product-like structure, but allows more flexible coefficients than AGP. It was first introduced in a quantum state-preparation setting and later implemented efficiently on a classical computer.\cite{Khamoshi2023JPCA,Dutta2024BTS} Previous work showed that BTS can improve substantially over AGP and gives promising results for model pairing and spin Hamiltonians, including the reduced BCS Hamiltonian and the one-dimensional XXZ model. In this way, BTS provides a promising variational approximation to seniority-zero physics while retaining polynomial computational cost.\cite{Dutta2024BTS}  

Recent work from our group introduced extended Jordan–Wigner Hartree–Fock (EJWHF) as another polynomial-cost variational alternative for the seniority-zero problem.\cite{JordanWigner1928,Wang1991,Henderson2022JCP,Chen2023JCP,HendersonGao2024,Henderson2024JCP,GhassemiTabrizi2025,Henderson2026JCTC} Here we study BTS in the same broader context. Our goal is not to present BTS as a universal replacement for pCCD or other seniority-zero methods. Rather, we examine it as another variational seniority-zero ansatz that offers a useful balance between flexibility, accuracy, and computational efficiency, while also being naturally connected to quantum state-preparation ideas.\cite{KhamoshiEvangelista2021,Khamoshi2023QST,Khamoshi2023JPCA,Dutta2024BTS} 

In this work, we apply BTS to the half-filled Hubbard model with both restricted and unrestricted orbital optimization. This is a useful setting because repulsive Hubbard physics favors singly occupied sites, making the quality of the orbital description especially important.\cite{Hubbard1963,WahlenStrothman2018,Henderson2026JCTC} In the unrestricted formulation, the spin-up and spin-down orbitals are optimized independently, and when expressed in the original site basis, the paired wave function can also represent configurations with electrons occupying different Hubbard sites. We compare BTS against Hartree–Fock, AGP, DOCI, and FCI where available. We also use BTS-optimized orbitals to test how closely they approach the orbital space favored by unrestricted DOCI through frozen-orbital energy comparisons. Finally, for small systems, we examine the onset of symmetry breaking through the Coulson–Fischer point.

\section{Theory}

\subsection{Hubbard Hamiltonian}
We consider the single-band Hubbard Hamiltonian in a general hopping-matrix form,\cite{Hubbard1963}
\begin{equation}
\hat{H}
=
\sum_{ij,\sigma}
t_{ij}
c_{i\sigma}^{\dagger}c_{j\sigma}
+
U\sum_i n_{i\uparrow}n_{i\downarrow},
\label{eq:hubbard-general}
\end{equation}
where $c_{i\sigma}^{\dagger}$ creates an electron with spin $\sigma$ on Hubbard site $i$,
$c_{i\sigma}$ annihilates an electron with spin $\sigma$ on site $i$, and
$n_{i\sigma}=c_{i\sigma}^{\dagger}c_{i\sigma}$ is the corresponding number operator.
The parameter $U$ is the on-site interaction strength, and $t_{ij}$ is the Hermitian hopping matrix which for the real nearest-neighbor Hubbard model will be
\begin{equation}
t_{ij}
=
\begin{cases}
-t, & i \text{ and } j \text{ are nearest-neighbor sites},\\
0, & \text{otherwise}.
\end{cases}
\label{eq:hopping-matrix}
\end{equation}We focus on half-filled lattices, where the number of electrons equals the number of lattice sites.
Throughout this work, $t$ sets the energy scale, and the correlation strength is controlled by $U/t$.

\subsubsection{Orbital rotation}

To optimize the orbitals, we use an \(S_z\)-preserving unitary orbital rotation, parameterized by the exponential of an anti-Hermitian one-particle generator.\cite{Thouless1960}

Let $\hat R$ denote the corresponding rotation operator. The rotation defines
a unitary transformation between the original Hubbard site operators,
$c_{p\sigma}^{\dagger}$, and the optimized orbital operators, $a_{r\sigma}^{\dagger}$,
\begin{align}
c_{p\sigma}^{\dagger}
&=
\sum_r R_{rp}^{\sigma} a_{r\sigma}^{\dagger},
\label{eq:orbital-rotation-creation}
\\
c_{p\sigma}
&=
\sum_r \left(R_{rp}^{\sigma}\right)^* a_{r\sigma}.
\label{eq:orbital-rotation-annihilation}
\end{align}
Here $R_{rp}^{\sigma}$ is a unitary one-particle orbital-rotation matrix for spin $\sigma$.
The rotation does not mix spin sectors: spin-up orbitals are rotated only among spin-up
orbitals, and spin-down orbitals are rotated only among spin-down orbitals.

Substituting Eqs.~\eqref{eq:orbital-rotation-creation} and
\eqref{eq:orbital-rotation-annihilation} into the Hubbard Hamiltonian gives the rotated Hamiltonian:
\begin{align}
\hat H_R
&=
\sum_{pqrs,\sigma}
t_{pq}R_{rp}^{\sigma}
\left(R_{sq}^{\sigma}\right)^*
a_{r\sigma}^{\dagger}a_{s\sigma}
\nonumber\\
&\quad+U
\sum_{p r s r' s'}
R_{rp}^{\uparrow}
\left(R_{sp}^{\uparrow}\right)^*
R_{r'p}^{\downarrow}
\left(R_{s'p}^{\downarrow}\right)^*
\nonumber\\
&\qquad\times
a_{r\uparrow}^{\dagger}a_{s\uparrow}
a_{r'\downarrow}^{\dagger}a_{s'\downarrow}.
\label{eq:rotated-hubbard}
\end{align}
Although the Hubbard interaction is local in the original site basis, the orbital rotation
generates a more general interaction in the optimized orbital basis. The next step is to
retain only the part of this rotated Hamiltonian that preserves seniority in the optimized
orbital basis.

\subsubsection{Seniority-zero part of the rotated Hamiltonian}
To retain the seniority-zero part of the rotated Hamiltonian in
Eq.~\eqref{eq:rotated-hubbard}, seniority is defined with respect to the optimized orbital basis.\cite{Bytautas2011,Henderson2026JCTC}

The pair creation, pair annihilation, and number operators in the optimized orbital basis are defined as:\cite{Surjan1999,TecmerBoguslawski2022}
\begin{equation}
P_r^{\dagger}
=
a_{r\uparrow}^{\dagger}a_{r\downarrow}^{\dagger},
\qquad
P_r
=
a_{r\downarrow}a_{r\uparrow},
\qquad
N_r
=
\sum_{\sigma}a_{r\sigma}^{\dagger}a_{r\sigma}.
\label{eq:pair-number-defs}
\end{equation}
Let $P_0$ denote the projector onto the seniority-zero sector. The one-body term survives the projection only when it acts within the paired spin orbitals:
\begin{equation}
P_0
\left(
\sum_{\sigma}a_{r\sigma}^{\dagger}a_{s\sigma}
\right)
P_0
=
\delta_{rs}N_r.
\label{eq:one-body-projection}
\end{equation}
Within the seniority-zero subspace, the two-body term gives two nonzero contributions:
\begin{align}
&P_0
\left(
a_{r\uparrow}^{\dagger}a_{s\uparrow}
a_{r'\downarrow}^{\dagger}a_{s'\downarrow}
\right)
P_0
\nonumber\\
&=
\delta_{rr'}\delta_{ss'}P_r^{\dagger}P_s
+
\delta_{rs}\delta_{s'r'}
\left(1-\delta_{rs'}\right)
\frac{1}{4}N_rN_{s'}.
\label{eq:two-body-projection}
\end{align}
The first term in Eq.~\eqref{eq:two-body-projection} moves an electron pair from orbital $s$ to orbital $r$ to define a pair-hopping, and the second term is a density--density interaction between two different spatial orbitals. In a more general local-seniority-conserving decomposition, the index pattern
\(r=s'\) and \(s=r'\) also gives an exchange-like spin term. For the seniority-zero
Hubbard Hamiltonian considered here, this term vanishes.

Using Eqs.~\eqref{eq:one-body-projection} and \eqref{eq:two-body-projection}, the
seniority-zero part of the rotated Hubbard Hamiltonian becomes
\begin{align}
P_0 \hat H_R P_0
&=
\sum_r \epsilon_r N_r
+U\sum_{rs}
\Biggl[
\left(
\sum_p C_{rs}^{(p)}
\right)
P_r^{\dagger}P_s
\nonumber\\
&\qquad
+
\frac{1}{4}
\left(
\sum_p D_{rs}^{(p)}
\right)
N_rN_s
\Biggr].
\label{eq:seniority-zero-rotated-hubbard}
\end{align}
Equation~\eqref{eq:seniority-zero-rotated-hubbard} is the effective Hamiltonian. For any seniority-zero trial state \( |\Psi\rangle \),
\(P_0|\Psi\rangle=|\Psi\rangle\).
An exact solution within this pairing space corresponds to DOCI, while BTS provides a variational approximation to that seniority-zero solution and For a common orbital basis, the variational energies satisfy \(E_{\mathrm{BTS}}\ge E_{\mathrm{DOCI}}\ge E_{\mathrm{FCI}}\).
The coefficients
\(\epsilon_r\), \(C_{rs}^{(p)}\), and \(D_{rs}^{(p)}\) depend on whether the orbital rotation is restricted or unrestricted, as defined below.

\subsubsection{Restricted and unrestricted representations}
The seniority-zero Hamiltonian in Eq.~\eqref{eq:seniority-zero-rotated-hubbard} has the
same operator structure for both restricted and unrestricted orbital rotations. The difference
is entirely in the orbital-rotation matrices entering the coefficients
\(\epsilon_r\), \(C_{rs}^{(p)}\), and \(D_{rs}^{(p)}\).
We first write the unrestricted form. In the unrestricted representation,
the spin-up and spin-down orbital rotations, \(R^\uparrow\) and
\(R^\downarrow\), are optimized independently and are not constrained
to be equal.
The one-body coefficient is
\begin{equation}
\epsilon_r
:=
\frac{1}{2}
\left(
\epsilon_r^{\uparrow}
+
\epsilon_r^{\downarrow}
\right),
\label{eq:unrestricted-epsilon}
\end{equation}
with
\begin{equation}
\epsilon_r^{\sigma}
:=
\sum_{pq}
t_{pq}
R_{rp}^{\sigma}
\left(R_{rq}^{\sigma}\right)^* .
\label{eq:epsilon-r-sigma}
\end{equation}
The pair-hopping coefficient is
\begin{equation}
C_{rs}^{(p)}
:=
\left(
R_{rp}^{\uparrow}R_{rp}^{\downarrow}
\right)
\left(
R_{sp}^{\uparrow}R_{sp}^{\downarrow}
\right)^* ,
\label{eq:unrestricted-C}
\end{equation}
and the density--density coefficient is
\begin{equation}
D_{rs}^{(p)}
:=
\left|R_{rp}^{\uparrow}\right|^2
\left|R_{sp}^{\downarrow}\right|^2
\left(1-\delta_{rs}\right).
\label{eq:unrestricted-D}
\end{equation}
The factor \((1-\delta_{rs})\) removes the same-orbital density--density contribution,
which is already included through the pair-hopping term. This additional freedom allows the variational wave function to break spin symmetry at the orbital level.

The restricted representation is obtained by imposing the same orbital rotation for both
spin sectors,
\begin{equation}
R^{\uparrow}=R^{\downarrow}\equiv R.
\label{eq:restricted-R}
\end{equation}
This distinction between unrestricted and restricted is central for the repulsive Hubbard model.\cite{Hubbard1963,WahlenStrothman2018}
The two representations differ in the spatial structure allowed for the spin-up and spin-down orbitals forming each pair. In the
restricted representation, the spin-up and spin-down members of a paired
level share the same spatial orbital, whereas in the unrestricted
representation their spatial components are optimized independently and
may differ. This additional flexibility permits spin-symmetry breaking
while retaining a seniority-zero paired description in the optimized basis.

When a paired level is expanded in the original Hubbard site basis, the
two electrons may occupy either the same site or different sites. A pair-creation operator associated with paired level \(p\) can be written as
\begin{align}
P_p^{\dagger}
&=
a_{p\uparrow}^{\dagger}a_{p\downarrow}^{\dagger}
\nonumber\\
&=
\sum_{ij}
\left(R_{pi}^{\uparrow}\right)^*
\left(R_{pj}^{\downarrow}\right)^*
c_{i\uparrow}^{\dagger}c_{j\downarrow}^{\dagger}.
\label{eq:pair-site-expansion}
\end{align}
The terms with \(i=j\) create doubly occupied Hubbard sites, whereas the terms with \(i\neq j\) place the two electrons on different sites.


\subsection{Binary tree state ansatz}

The BTS ansatz can be viewed as a direct generalization of the coefficient structure of AGP.\cite{Dutta2024BTS} In the optimized orbital basis, the AGP wave function is written as\cite{Coleman1965,BratozDurand1965,Linderberg1980}

\begin{align}
\ket{\AGP}
&=
\sum_{1\leq p_1<\cdots<p_n\leq m}
\eta_{p_1}\eta_{p_2}\cdots\eta_{p_n}
\nonumber\\
&\quad\times
P_{p_1}^{\dagger}
P_{p_2}^{\dagger}
\cdots
P_{p_n}^{\dagger}
\ket{-} .
\label{eq:agp-ordered}
\end{align}

BTS generalizes AGP by allowing the geminal amplitudes to depend
on the pair index.\cite{Dutta2024BTS,Khamoshi2023JPCA} For \(n\) electron pairs distributed among \(m\) optimized spatial orbitals, the BTS
wave function is written as
\begin{align}
\ket{\BTS}
&=
\sum_{1\leq p_1<\cdots<p_n\leq m}
\eta_{p_1}^{1}
\eta_{p_2}^{2}
\cdots
\eta_{p_n}^{n}
\nonumber\\
&\quad\times
P_{p_1}^{\dagger}
P_{p_2}^{\dagger}
\cdots
P_{p_n}^{\dagger}
\ket{-}.
\label{eq:bts-wavefunction}
\end{align}
where \(P_p^\dagger\) creates an electron pair in optimized orbital \(p\), and
\(\eta_p^j\) is the BTS amplitude associated with placing the \(j\)-th pair in
orbital \(p\). Note that \(j\) is an index and not an exponent. The paired-level labels follow the ordering of the orbital-rotation matrices. In the unrestricted representation, the spin-up and spin-down orbitals carry the same level index \(p\).

The BTS structure comes from the binary-tree polynomial (BTP) which for scalar quantities \(A_p^j\) defines
\begin{align}
B_n^m(\mathbf{A})
&=
\sum_{1\leq p_1<\cdots<p_n\leq m}
A_{p_1}^{1}
A_{p_2}^{2}
\cdots
A_{p_n}^{n}.
\label{eq:btp}
\end{align}
The BTS wave function is obtained by replacing the scalar entries by pair-creation operators,
\begin{equation}
\ket{\BTS}
=
B_n^m(\mathbf{A})\ket{-},
\qquad
A_p^j
=
\eta_p^j P_p^\dagger.
\label{eq:bts-as-btp}
\end{equation}
The BTP also satisfies the recursion relation
\begin{equation}
B_j^p(\mathbf{A})
=
A_p^j B_{j-1}^{p-1}(\mathbf{A})
+
B_j^{p-1}(\mathbf{A}),
\label{eq:btp-recursion}
\end{equation}
\begin{equation}
B_0^p(\mathbf{A})=1,
\qquad
B_j^p(\mathbf{A})=0
\quad
\text{for } j<0 \text{ or } j>p .
\label{eq:btp-boundary}
\end{equation}
The first term in Eq.~\eqref{eq:btp-recursion} includes orbital \(p\) as the location of
the \(j\)-th pair, while the second term excludes orbital \(p\). This recursion gives the ansatz its binary-tree structure.

The same recursive structure is used to evaluate BTS overlaps and reduced density matrices efficiently, as described in the original BTS formulation.\cite{Dutta2024BTS}

For fixed orbitals, the binary-tree polynomial overlaps can be evaluated in \(O(mn)\) time, and the corresponding single-BTS variational energy can be evaluated with \(O(m^4)\) scaling in the number of paired orbitals. Here \(m\) is the number of spatial orbitals, or paired levels, and \(n\) is the number of electron pairs.\cite{Dutta2024BTS}  This should be contrasted with the \(O(m^3)\) cost of pCCD\cite{Henderson2014} and the \(O(m^5)\) cost of EJW-HF (this cost arising from the \(O(m^2)\) Hamiltonian terms that carry JW strings, each of which can be evaluated in \(O(m^3)\) cost by the construction of a transition density matrix).\cite{Chen2023JCP, Henderson2026JCTC}

\section{Results}

To assess the performance of BTS, unless otherwise stated, we consider half-filled Hubbard lattices with open boundary conditions. 
All BTS calculations are carried out with the seniority-zero part of the rotated Hubbard Hamiltonian discussed above. 
For each value of \(U/t\), both the BTS amplitudes and the orbital-rotation
parameters are optimized variationally using L-BFGS with a backtracking
line search. Orbital derivatives are evaluated by automatic differentiation,
while derivatives with respect to the BTS amplitudes are evaluated analytically
using transition reduced density matrices. The gradient tolerance is
\(10^{-8}\), followed by a tight polish to \(10^{-11}\). The \(U/t\) scans
are warm-started from neighboring interaction strengths and supplemented by
randomized orbital and BTS-amplitude starting guesses. The DOCI calculations were performed using an in-house code.

\begin{figure}[h!]
\centering
\includegraphics[width=\columnwidth]{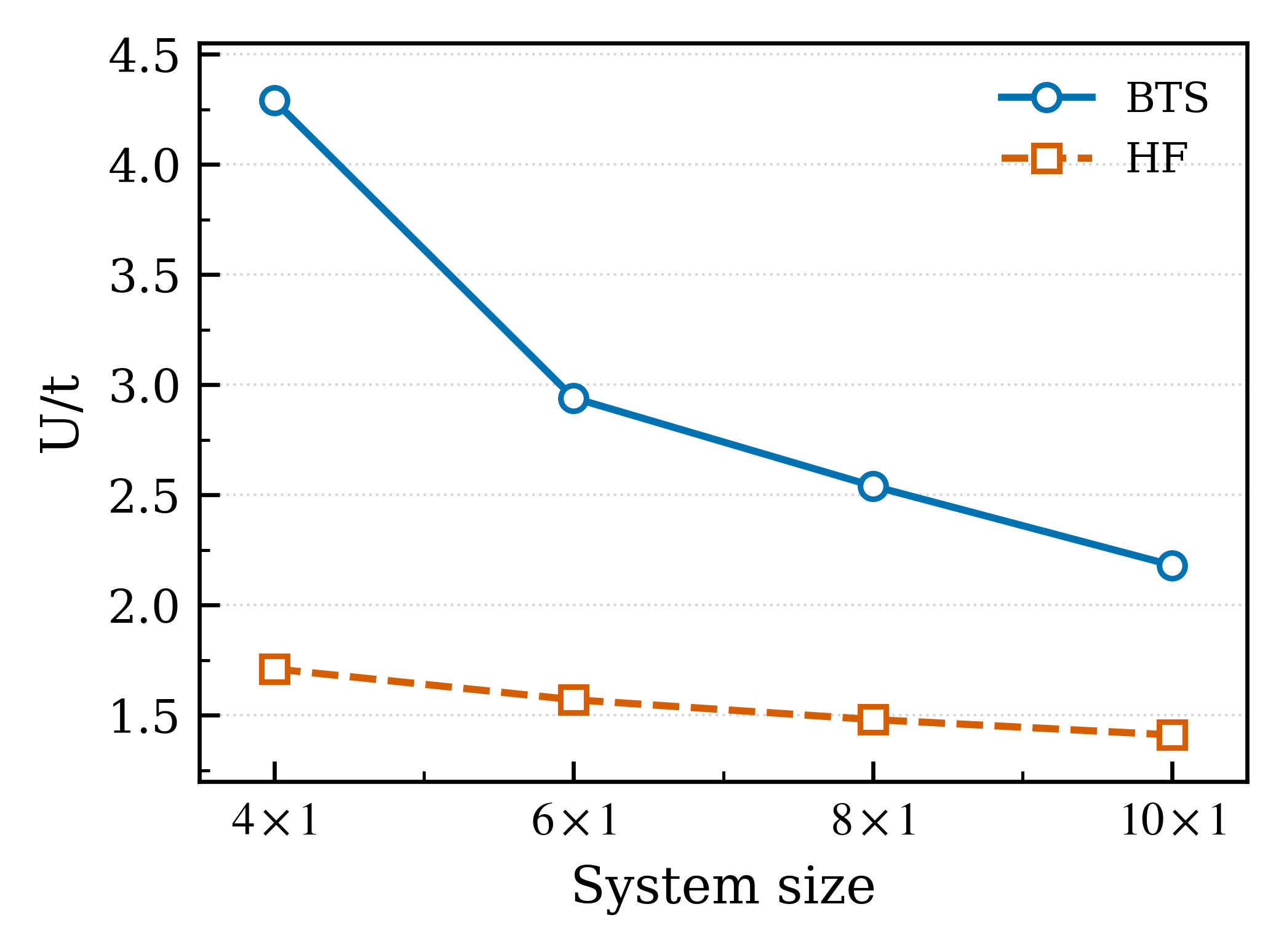}
\caption{Coulson--Fischer point for BTS and Hartree--Fock as a function of one-dimensional Hubbard lattice size.}
\label{fig:cf-point}
\end{figure}

Before discussing the unrestricted BTS energies, we first examine the Coulson--Fischer point as a function of system size. Spin-symmetry
breaking is monitored through the expectation value
\(\langle \hat S^2\rangle\). The onset is identified by the
sharp departure of \(\langle \hat S^2\rangle\) from its numerical
zero value to a finite broken-symmetry value, using a grid spacing of
\(\Delta(U/t)=0.01\).

As shown in Fig.~\ref{fig:cf-point}, the BTS Coulson--Fischer point occurs at larger \(U/t\) than the Hartree--Fock one for all system sizes considered. This indicates that restricted BTS can preserve the restricted symmetry over a wider interaction range than restricted HF. 
In restricted HF, the wave function has limited variational freedom to describe the localization induced by the repulsive Hubbard interaction. As a result, HF lowers its energy by breaking spin symmetry at relatively small \(U/t\). In contrast, the correlated pair structure in BTS delays the need for symmetry breaking as a mechanism for lowering the energy.

For both BTS and HF, the Coulson--Fischer onset decreases with system
size, showing that the unrestricted variational solutions begin to break
spin symmetry at smaller \(U/t\) as the lattice grows.

\begin{figure*}[t!]
\centering
\includegraphics[width=0.98\textwidth]{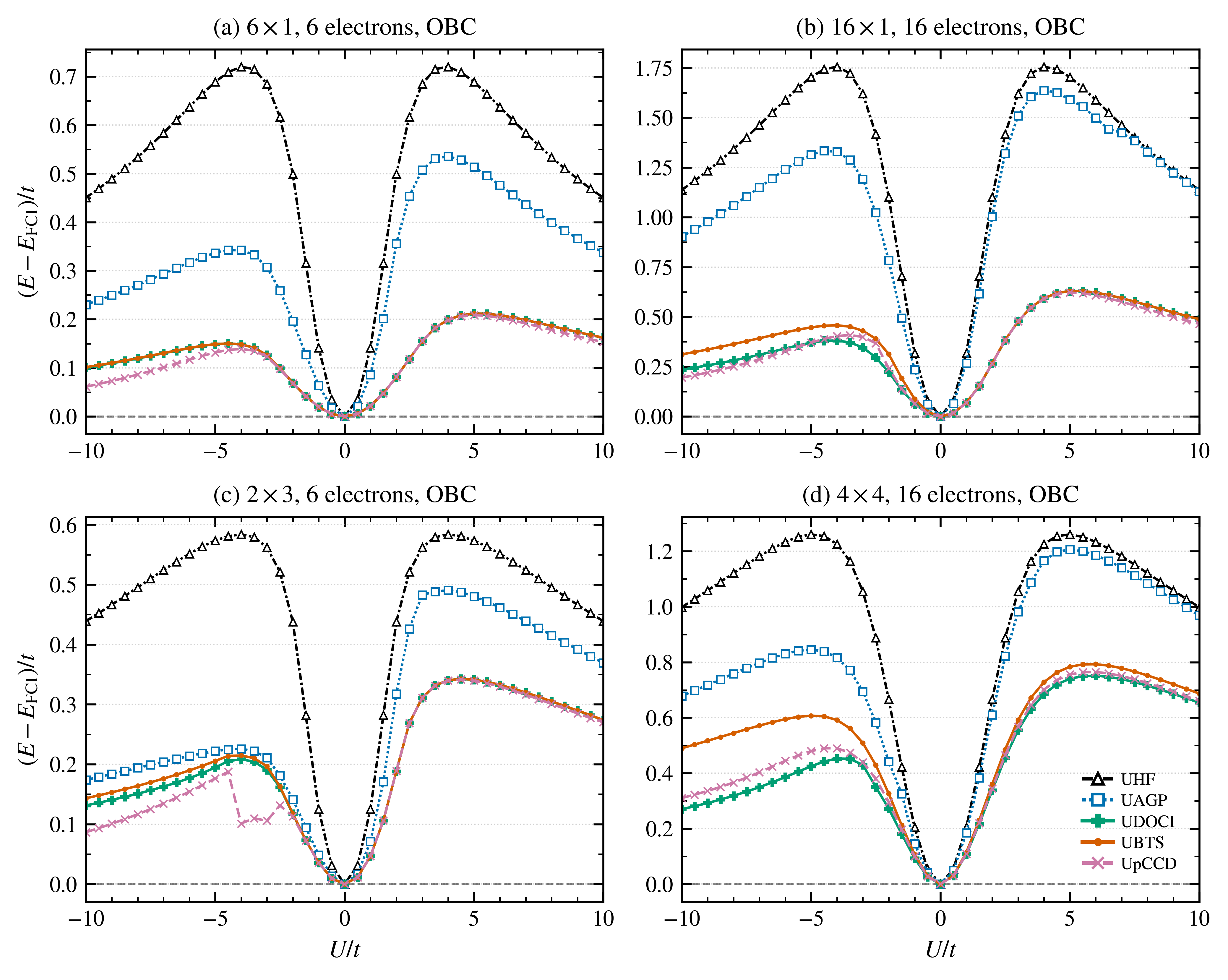}
\caption{Total energy errors relative to FCI for half-filled Hubbard lattices
with open boundary conditions: (a) \(6\times1\), (b) \(16\times1\),
(c) \(2\times3\), and (d) \(4\times4\).
The pCCD calculations use frozen UBTS-optimized orbitals.
The legend in panel (d) applies to all panels.}
\label{fig:energy-errors}
\end{figure*}

We begin studying the energetic accuracy of BTS with one-dimensional half-filled Hubbard chains with open boundary conditions, shown in the
top row of Fig.~\ref{fig:energy-errors}. 
The plotted quantity is the error with respect to FCI, allowing HF, AGP, DOCI, and BTS to be compared on the same scale. 
On the repulsive side, unrestricted BTS follows unrestricted DOCI very closely. 
The maximum UBTS error relative to UDOCI is \(3.5\times 10^{-5}t\) per site for the \(6\times 1\) chain and \(1.1\times 10^{-4}t\) per site for the \(16\times 1\) chain over the range \(0\leq U/t\leq 10\). 
This shows that, in one dimension, BTS provides a compact variational approximation to the optimized seniority-zero solution. The DOCI--FCI separation in Fig.~\ref{fig:energy-errors} reflects contributions outside the seniority-zero space.

The comparison with AGP is also informative. The AGP parameters and orbitals are likewise optimized with L-BFGS using a
gradient tolerance of \(10^{-8}\), with randomized orbital restarts used to
locate the lowest converged solution.
Both AGP and BTS are geminal-based paired wave functions, but BTS has a more flexible coefficient structure. 
This additional flexibility is especially visible on the repulsive side, where BTS remains close to DOCI while AGP has a substantially larger error. 
The attractive side is less diagnostic for unrestricted orbitals. 
For \(U<0\), the Hubbard interaction favors pairing and double occupancy, so restricted and unrestricted paired wave functions tend to give similar solutions. 
Nevertheless, BTS remains variational and the orbital optimization remains stable in cases where unrestricted AGP can become difficult to converge for larger systems and stronger interactions.

The bottom row of Fig.~\ref{fig:energy-errors} shows the corresponding results for two-dimensional lattices. 
As expected, the two-dimensional systems are more challenging. 
The \(2\times 3\) lattice is still described quite accurately by BTS relative to UDOCI on the repulsive side, with a maximum UBTS error relative to UDOCI of \(1.96\times 10^{-4}t\) per site for \(0\leq U/t\leq 10\). 
For the \(4\times 4\) lattice, the error is larger, reaching \(2.71\times 10^{-3}t\) per site near intermediate coupling. 
Even in this more difficult case, however, BTS gives a clear improvement over AGP and remains much closer to DOCI in the repulsive regime. 
This suggests that unrestricted BTS captures a significant part of the orbital relaxation and correlation effects needed in the repulsive Hubbard model, although the agreement with DOCI becomes less quantitative as the dimensionality and system size increase.

\begin{figure}[t!]
\centering
\includegraphics[width=\columnwidth]{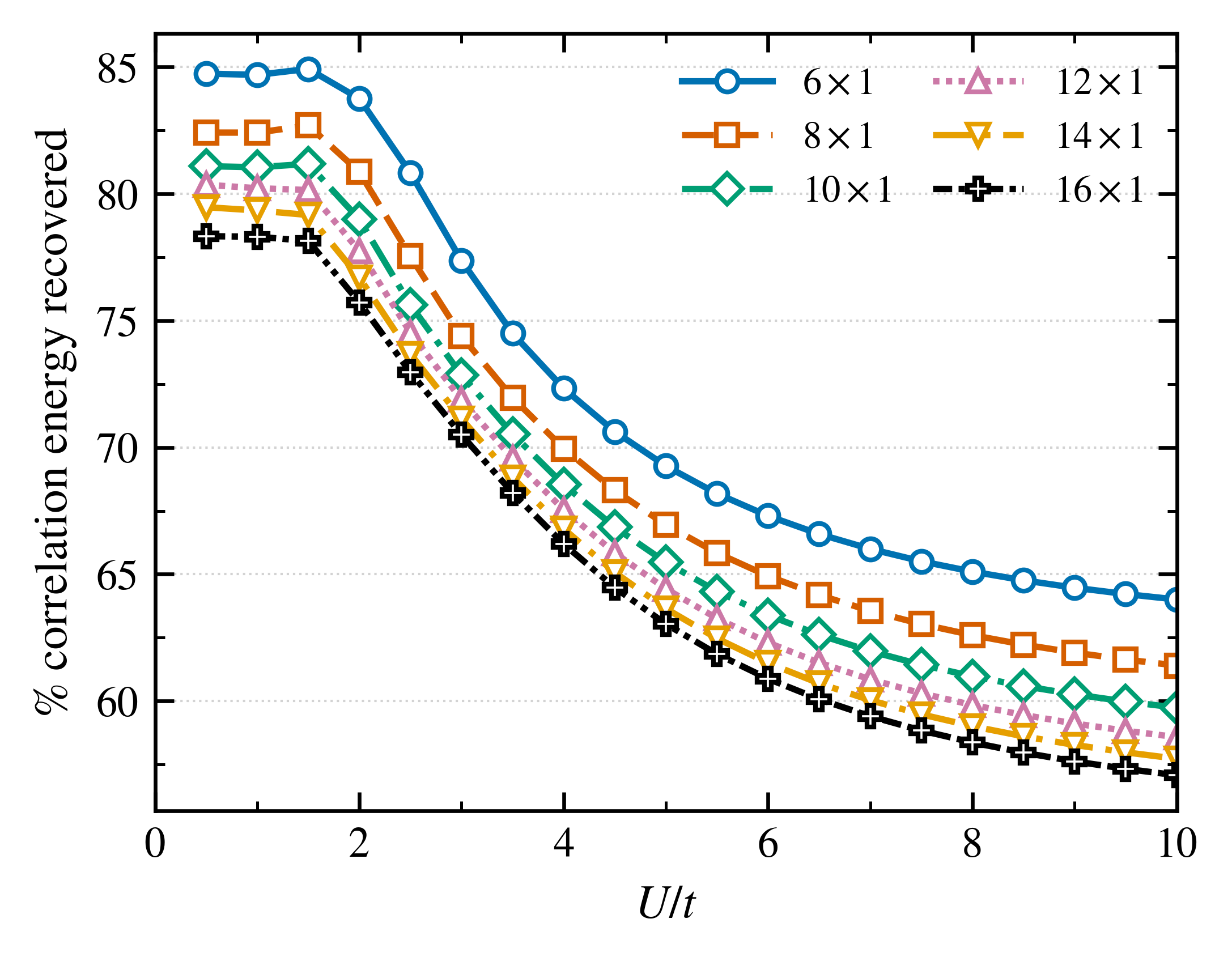}
\caption{Percentage of FCI correlation energy recovered by UBTS for repulsive half-filled one-dimensional Hubbard chains with open boundary conditions. The percentage is defined relative to UHF and FCI according to Eq.~\eqref{eq:corr-recovered}.}
\label{fig:corr-recovered}
\end{figure}

To focus more directly on the repulsive regime, Fig.~\ref{fig:corr-recovered} shows the percentage of FCI correlation energy recovered by UBTS in one-dimensional chains. 
We define this quantity as
\begin{equation}
\%E_{\mathrm{corr}}
=
100\times
\frac{
E_{\mathrm{UHF}} - E_{\mathrm{UBTS}}
}{
E_{\mathrm{UHF}} - E_{\mathrm{FCI}}
}.
\label{eq:corr-recovered}
\end{equation}
With this definition, \(0\%\) corresponds to UHF and \(100\%\) corresponds to FCI. 
Across the system sizes considered, UBTS recovers a sizeable fraction of the total correlation energy before any additional correlation treatment is applied. 
The percentage recovered is largest at weaker and intermediate interactions, and decreases as \(U/t\) becomes large. 
This trend is physically reasonable: at large repulsive \(U/t\), the Hubbard model develops stronger singly occupied-site character, and a seniority-zero paired ansatz remains a restricted representation of the full Hilbert space. 
Even so, UBTS recovers \(50\%\) or more of the correlation energy over much of the repulsive range, showing that the variational BTS reference captures a substantial fraction of the total correlation energy.

Except for pCCD, for which the UBTS-optimized orbitals are held fixed, the energies discussed above are obtained by optimizing the wave-function parameters and orbital rotations for each method.
To separate the quality of the BTS orbitals from the quality of the BTS coefficient parameterization, we also perform frozen-orbital UDOCI calculations using the orbitals optimized by UBTS. 
The orbital error is defined as
\begin{equation}
\Delta E/N
=
\frac{
E_{\mathrm{UDOCI}}(O_{\mathrm{UBTS}})
-
E_{\mathrm{UDOCI}}(O_{\mathrm{UDOCI}})
}{N},
\label{eq:orbital-error}
\end{equation}
where \(O_{\mathrm{UBTS}}\) denotes the UBTS-optimized orbital rotation and \(O_{\mathrm{UDOCI}}\) denotes the fully optimized UDOCI orbital rotation.

\begin{figure}[t!]
\centering
\includegraphics[width=\columnwidth]{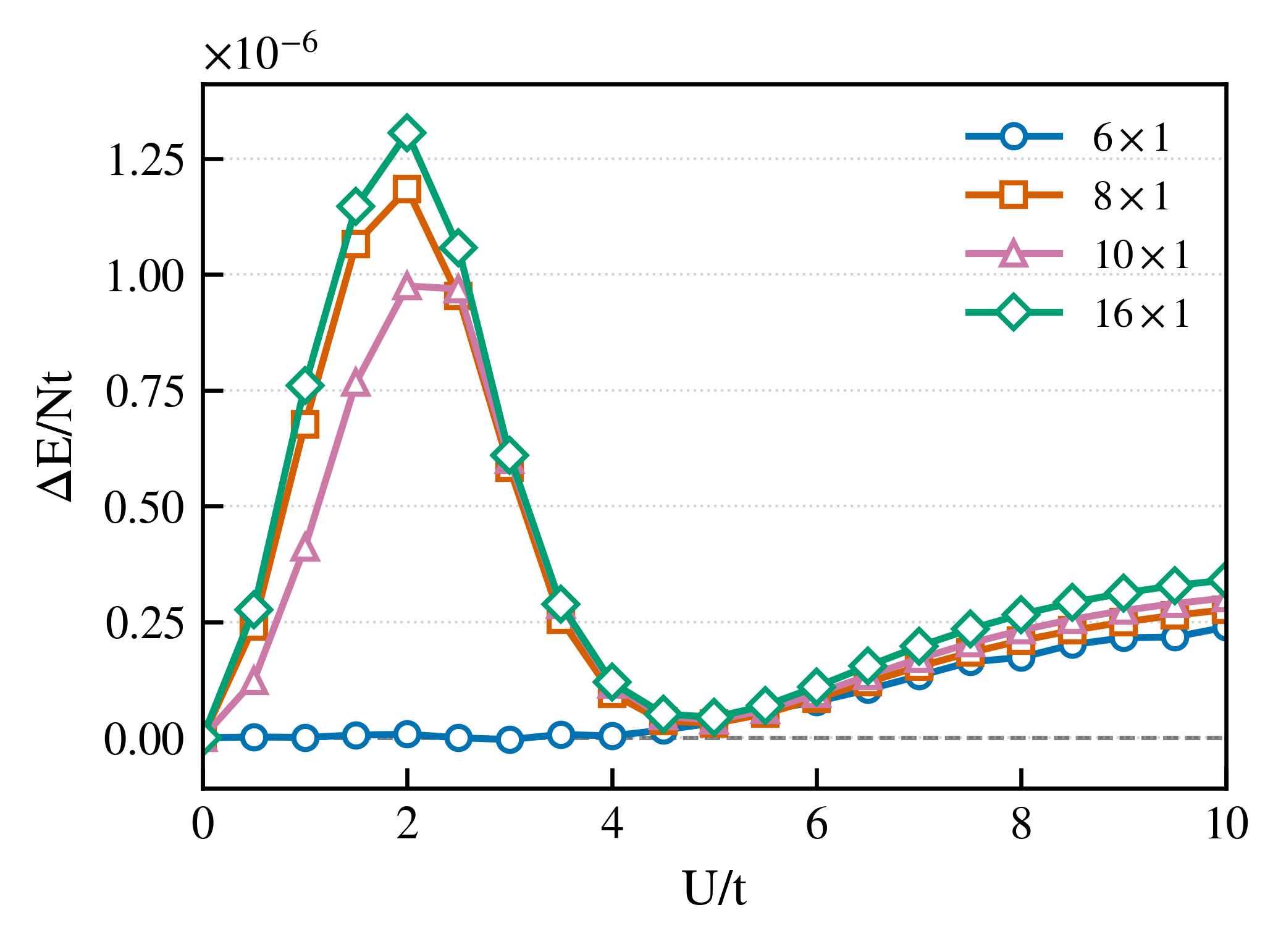}
\caption{Frozen-orbital UDOCI error obtained using UBTS-optimized orbitals for repulsive half-filled one-dimensional Hubbard chains with open boundary condition.}
\label{fig:orbital-error}
\end{figure}

The results are shown in Fig.~\ref{fig:orbital-error}. 
For the one-dimensional chains, the frozen-orbital UDOCI error remains on the order of \(10^{-6}t\) per site over the repulsive range. 
Thus, the orbitals obtained from UBTS are very close, in an energetic sense, to the orbitals preferred by UDOCI. 
The largest frozen-orbital energy errors occur at small positive \(U/t\). At larger \(U/t\), the error remains small and varies smoothly with system size.
This indicates that the dominant remaining difference between UBTS and UDOCI is not primarily due to poor orbital optimization, but rather to the finite flexibility of the BTS coefficient structure.

For comparison, Fig.~\ref{fig:energy-errors} also includes unrestricted
pCCD calculations evaluated using the UBTS-optimized orbitals.
The occupied paired levels defining the pCCD reference are selected from
the BTS number occupations, with the \(n\) levels
of the highest occupation assigned as occupied, and the remaining levels as
virtual. The pCCD cluster operator is
\begin{equation}
T_2=\sum_{ia} t_i^a P_a^\dagger P_i,
\end{equation}
where \(i\) and \(a\) label occupied and virtual paired levels,
respectively. The amplitudes are obtained from the projected pCCD
equations at fixed UBTS orbitals, with the maximum amplitude residual
converged below \(10^{-9}\).
This comparison is particularly interesting on the attractive side.
pCCD is known to perform poorly in strongly attractive pairing regimes,
where it may overcorrelate or even break down.\cite{Henderson2026JCTC}
For the attractive Hubbard solutions considered here, the restricted and unrestricted
paired solutions give the same energies, so the use of unrestricted pCCD does not by
itself remove this known difficulty.
Nevertheless, when the UBTS-optimized orbitals are used as a frozen orbital basis,
pCCD gives surprisingly reasonable energies throughout much of the attractive regime.
This occurs even though the UBTS orbitals themselves are not as close to the
DOCI-optimal orbitals on the attractive side as they are for repulsive \(U\).

The pCCD solutions remain well converged across the nonsmooth region
of the attractive \(2\times3\) lattice. The feature coincides with changes
in the frozen UBTS orbital basis, reflecting the sensitivity of pCCD to
the orbital representation used for the reference.

We note, however, that good agreement with DOCI energies does not necessarily imply equally accurate reduced density matrices. We see this by considering the opposite spin density-density correlation function, which we define as 
\begin{equation}
C_{\uparrow\downarrow}(r)
=
\frac{1}{L}
\sum_i
\frac{
\bra{\Psi}
n_{i\uparrow}n_{i+r,\downarrow}
\ket{\Psi}
}{
\braket{\Psi|\Psi}
},
\label{eq:site-density-correlation}
\end{equation}
where \(L\) is the number of lattice sites and, for periodic boundary conditions, the displaced site is
\(j=1+[(i+r-1)\bmod L]\).
For the half-filled periodic chain, we report \(r=0,\ldots,L/2\).

Because the wave functions are defined in the optimized pairing basis, the site-basis operator
\(n_{i\uparrow}n_{j\downarrow}\) is evaluated by transforming it into the optimized orbital basis.
Equivalently, it can be written in terms of the pair-transfer and density-density reduced density matrices as
\begin{align}
\langle n_{i\uparrow}n_{j\downarrow}\rangle
&=
\frac{1}{\langle\Psi|\Psi\rangle}
\sum_{pq}
R_{pi}^{\uparrow}
\left(R_{qi}^{\uparrow}\right)^*
R_{pj}^{\downarrow}
\left(R_{qj}^{\downarrow}\right)^*
\langle P_p^\dagger P_q\rangle
\nonumber\\
&\quad+
\frac{1}{4\langle\Psi|\Psi\rangle}
\sum_{p\neq q}
\left|R_{pi}^{\uparrow}\right|^2
\left|R_{qj}^{\downarrow}\right|^2
\langle N_pN_q\rangle .
\label{eq:site-density-rdm}
\end{align}
The precise placement of complex conjugates follows the orbital-rotation convention in Eq.~\eqref{eq:orbital-rotation-creation}.

For BTS and DOCI, the expectation values are evaluated with the
normalized Hermitian wave function. For pCCD, the corresponding density
matrices are evaluated using the conventional biorthogonal coupled-cluster
expectation value
\begin{equation}
\langle O\rangle_{\mathrm{pCCD}}
=
\langle\Phi_0|(1+\Lambda)e^{-T_2} O e^{T_2}|\Phi_0\rangle
\end{equation}
where the \(\Lambda\) amplitudes are obtained from the projected
left-state equations.

\begin{figure}
\centering
\includegraphics[width=\columnwidth]{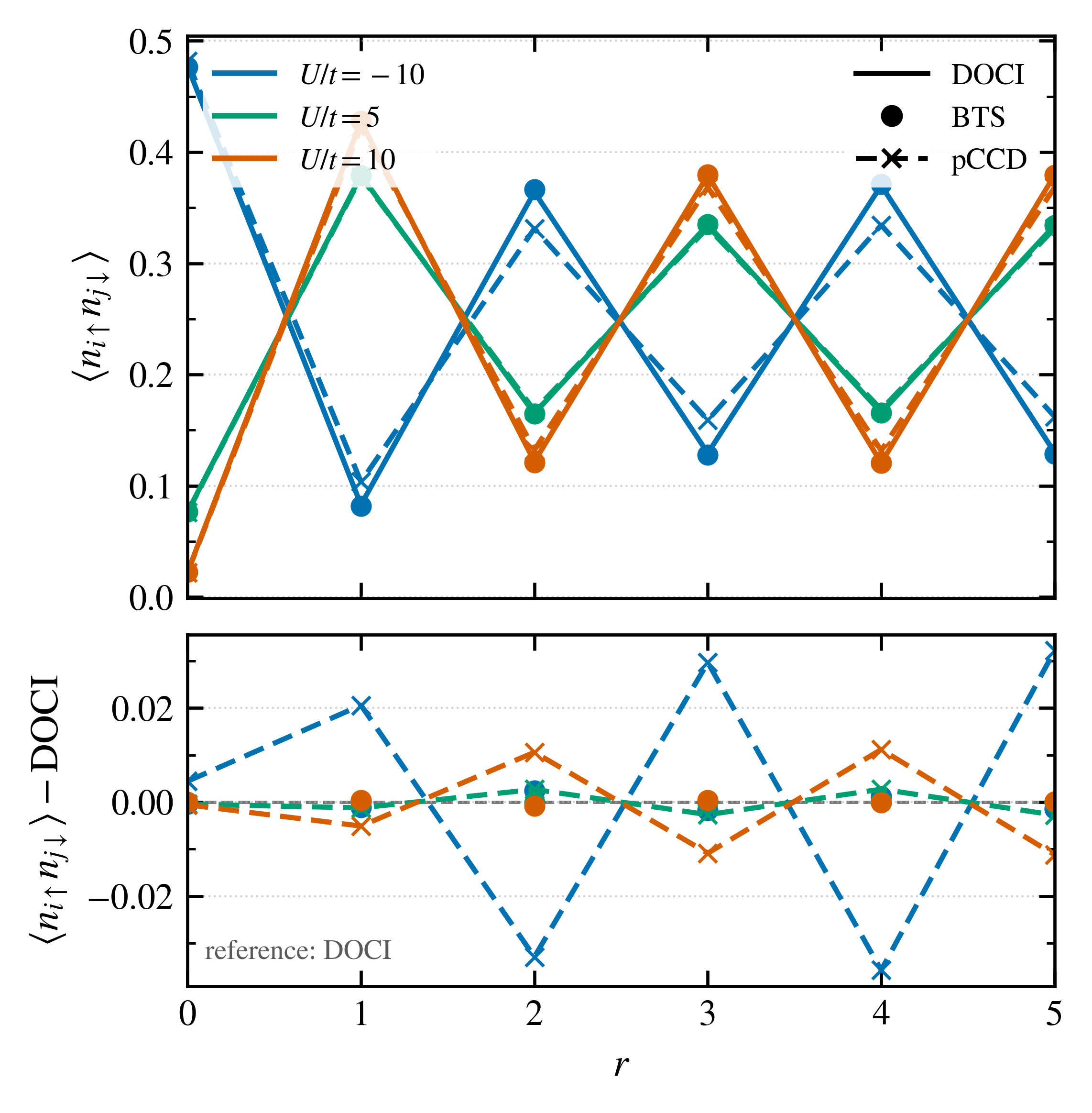}
\caption{
Site-basis opposite-spin density correlation function
for the \(10\times 1\) half-filled periodic Hubbard chain.
All methods are evaluated using the same UBTS-optimized orbitals.
The top panel compares DOCI, BTS, and pCCD, while the bottom panel shows the deviation from DOCI.
}
\label{fig:density-correlation}
\end{figure}

Figure~\ref{fig:density-correlation} shows \(C_{\uparrow\downarrow}(r)\) for the \(10\times 1\) half-filled periodic Hubbard chain.
To isolate the quality of the many-body ansatz from the quality of the orbital optimization, all three methods are evaluated using the same UBTS-optimized orbitals.
The DOCI result in this orbital basis is used as the reference.

The repulsive cases show the expected antiferromagnetic pattern.
For \(U/t=5\) and \(U/t=10\), the opposite-spin density correlation alternates strongly with distance: opposite-spin occupation is enhanced on neighboring sites and suppressed on the same sublattice.
At \(r=0\), the correlation function is small for \(U/t=10\), reflecting the suppression of double occupancy by the repulsive Hubbard interaction.
In contrast, the attractive case has a large on-site value \(C_{\uparrow\downarrow}(0)\), consistent with the attractive interaction favoring local up-down pairing.

The lower panel shows the deviation from DOCI in the same UBTS orbital basis.
The pCCD correlation function deviates substantially more from DOCI than BTS does.
This is especially clear on the attractive side.
This behavior is consistent with previous observations that pCCD, although often very accurate for DOCI energies, can be much less reliable for DOCI density matrices.\cite{Henderson2026JCTC}
Since the site-basis correlation function in Eq.~\eqref{eq:site-density-rdm} is constructed from pair-transfer and density--density reduced density matrices, errors in the pCCD density matrices translate directly into errors in the correlation function.
In contrast, BTS reproduces the DOCI
opposite-spin density correlation function very closely for all three
interaction strengths shown in the common UBTS-optimized orbital basis.

\section{Conclusions}
In this work, we applied restricted and unrestricted orbital-optimized binary tree state wave functions to the half-filled Hubbard model. The ansatz remains seniority zero in the optimized pairing basis, while unrestricted orbital optimization gives the two spin sectors independent spatial flexibility. In the Hubbard site basis, this paired description can accommodate the singly occupied-site configurations that become important in the repulsive regime. This makes unrestricted BTS a natural paired ansatz for the repulsive Hubbard model, where double occupancy is suppressed and spin symmetry breaking becomes energetically important.

The results show that BTS provides a useful variational approximation to the optimized seniority-zero solution. In one-dimensional open-boundary Hubbard chains, UBTS closely tracks UDOCI on the repulsive side, with errors remaining small across the range of \(U/t\) considered. The agreement is less quantitative in two dimensions, especially for the \(4\times 4\) lattice, but BTS still gives a clear improvement over AGP and remains much closer to DOCI in the repulsive regime. The Coulson--Fischer analysis shows that BTS breaks spin symmetry later than Hartree--Fock, indicating that the correlated pair structure of the restricted BTS wave function delays the need for unrestricted orbital relaxation.

We also examined the quality of the orbitals obtained from UBTS. Frozen-orbital UDOCI calculations using the UBTS-optimized orbitals give very small energy errors in the one-dimensional repulsive systems. This suggests that UBTS does not merely give a reasonable energy; it also identifies an orbital basis close to, in an energetic sense, the one preferred by UDOCI. The remaining difference between UBTS and UDOCI therefore appears to come mainly from the finite flexibility of the BTS coefficient structure rather than from a failure of the orbital optimization.

From a computational perspective, BTS offers a useful compromise between accuracy and cost.
The original formulation of BTS for classical computers evaluates the required reduced density matrices with quartic scaling in the number of paired orbitals, and the total cost of orbital optimization depends on the orbital-gradient implementation and the number of optimization steps.
Nevertheless, the underlying BTS energy expression remains fully variational.
This distinction is important when comparing BTS with pCCD.
Although pCCD is an efficient approximation to DOCI, it is nonvariational, and its orbital optimization must be handled with care, especially in unrestricted case.
Moreover, pCCD is known to reproduce DOCI energies more reliably than DOCI density matrices.
This limitation is reflected in the site-basis density correlations reported here: although all methods are evaluated in the same UBTS-optimized orbital basis, pCCD shows noticeably larger deviations from DOCI, especially on the attractive side.
In contrast, BTS closely reproduces the DOCI opposite-spin density correlation function, suggesting that its reduced density matrices are also reliable for this test.
BTS therefore provides a different polynomial-cost seniority-zero parameterization: it is variational, contains AGP as a special case, and is more flexible than AGP while remaining suitable as a compact reference wave function.

The connection of BTS to quantum state preparation\cite{Khamoshi2023JPCA} also makes it attractive as a starting point for future quantum and hybrid quantum-classical implementations. The present results suggest that BTS can serve as a practical variational reference for strongly correlated lattice Hamiltonians, particularly on the repulsive side of the Hubbard model. Future work should explore post-BTS correlation, linear combinations of binary tree states, and extensions that include higher-seniority or dynamic correlation effects beyond the seniority-zero sector.

\section*{Data Availability}
The data that support the findings of this study are available from the corresponding author upon reasonable request.

\begin{acknowledgments}
This work was supported by the U.S. Department of Energy, Office of Basic Energy Sciences, under Award DE-SC0001474.  G.E.S. is a Welch Foundation Chair (C-0036). We also thank Dr. Rishab Dutta for developing the original Julia implementation of BTS.
\end{acknowledgments}

\bibliography{ref}

\end{document}